# Structure, spin correlations and magnetism of the $S$ = 1/2 square-lattice antiferromagnet $Sr_2CuTe_{1-x}W_xO_6$ ($0 \leq x \leq 1$)

Otto H. J. Mustonen,[1,2]* Ellen Fogh,[3]* Joseph A. M. Paddison,[4] Lucile Mangin-Thro,[5] Thomas Hansen,[5] Helen Playford,[6] Maria Diaz-Lopez,[7] Peter Babkevich,[3] Sami Vasala,[8] Maarit Karppinen,[9] Edmund J. Cussen,[2] Henrik M. Rønnow,[3] Helen C. Walker[6]*

1 School of Chemistry, University of Birmingham, Birmingham B15 2TT, United Kingdom
2 Department of Material Science and Engineering, University of Sheffield, Sheffield S1 3JD, United Kingdom
3 Laboratory for Quantum Magnetism, Institute of Physics, École Polytechnique Fédérale de Lausanne (EPFL), CH-1015 Lausanne, Switzerland
4 Materials Science and Technology Division, Oak Ridge National Laboratory, Oak Ridge, Tennessee 37831, USA
5 Institut Laue Langevin, 71 Avenue des Martyrs, CS 20156, F-38042 Grenoble Cedex 9, France
6 ISIS Neutron and Muon Source, Rutherford Appleton Laboratory, Chilton, Didcot OX11 OQX, United Kingdom
7 Université Grenoble Alpes, CNRS, Grenoble INP, Institut Néel, 38000 Grenoble, France
8 ESRF - The European Synchrotron, 38000 Grenoble, France
9 Department of Chemistry and Materials Science, Aalto University, FI-00076, Espoo, Finland

*Corresponding authors:
o.mustonen@bham.ac.uk, ellen.fogh@epfl.ch, helen.c.walker@stfc.ac.uk

Quantum spin liquids are highly entangled magnetic states with exotic properties. The $S$ = 1/2 square-lattice Heisenberg model is one of the foundational models in frustrated magnetism with a predicted, but never observed, quantum spin liquid state. Isostructural double perovskites $Sr_2CuTeO_6$ and $Sr_2CuWO_6$ are physical realizations of this model, but have distinctly different types magnetic order and interactions due to a $d^{10}/d^0$ effect. Long-range magnetic order is suppressed in the solid solution $Sr_2CuTe_{1-x}W_xO_6$ in a wide region of $x$ = 0.05–0.6, where the ground state has been proposed to be a disorder-induced spin liquid. Here we show that the spin-liquid-like $x$ = 0.2 and $x$ = 0.5 samples have distinctly different local spin correlations, which suggests they have different ground states. Furthermore, the previously ignored interlayer coupling between the square-planes is likely to play a role in the suppression of magnetic order on the W-rich side at $x \approx 0.6$. These results highlight the complex magnetism of $Sr_2CuTe_{1-x}W_xO_6$ and hint at a new quantum critical point at $x \approx 0.3$.

**INTRODUCTION**

Spin-1/2 square-lattice antiferromagnets have been of significant scientific interest since the discovery of high-$T_c$ superconductivity in cuprates.[1] These antiferromagnetic and insulating parent phases become superconducting upon hole or electron doping.[2] The $S$ = 1/2 square-lattice Heisenberg model is also one of the foundational models of frustrated magnetism.[3] This model has two magnetic interactions: the nearest-neighbor $J_1$ interaction along the side of the square and the next-nearest-neighbor $J_2$ along the diagonal. A dominant antiferromagnetic $J_1$ leads to Néel antiferromagnetic order as observed in the high-$T_c$ parent phases, while dominant $J_2$ leads to columnar antiferromagnetic order. The competition between antiferromagnetic $J_1$ and $J_2$ interactions leads to magnetic frustration, which is predicted to stabilize a quantum spin liquid in the highly frustrated $J_2/J_1 \approx$ 0.4–0.6 region.[3–10] Quantum spin liquids are exotic quantum states consisting of highly entangled spins, that remain dynamic even at absolute zero without magnetic order or spin freezing.[11–13] A number of $S$ = 1/2 square-lattice antiferromagnets are known with either Néel or columnar antiferromagnetic order,[14–20] but the predicted quantum spin liquid state has never been observed. However, recent theoretical studies propose disorder as a possible route for stabilising a spin-liquid-like state.[21–25]

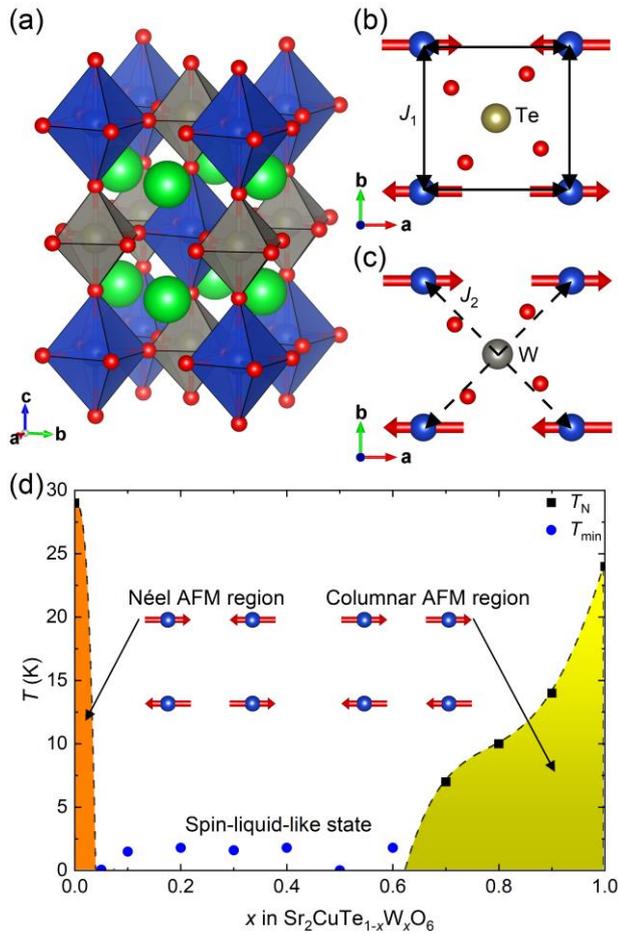

Figure 1. (a) The *B*-site ordered double perovskite structure of $Sr_2CuTe_{1-x}W_xO_6$, where the green, blue, gray and red spheres represent Sr, Cu, Te/W and O, respectively. The magnetic interactions are highly two-dimensional in *ab* plane, where the $S = 1/2$ $Cu^{2+}$ cations form a square lattice. (b) The Néel antiferromagnetic structure of $Sr_2CuTeO_6$ ($x = 0$) stabilised by a dominant antiferromagnetic $J_1$ interaction. (c) The columnar magnetic structure of $Sr_2CuWO_6$ ($x = 1$) stabilised by a dominant antiferromagnetic $J_2$ interaction. (d) Magnetic phase diagram of the solid solution $Sr_2CuTe_{1-x}W_xO_6$ based on muon spectroscopy from refs. [48,50]. The black squares represent measured Néel temperatures and the blue circles represent the lowest temperatures measured, where the magnetism remains dynamic. A spin-liquid-like state without magnetic order is observed between $x = 0.05$ and $x = 0.6$.

The *B*-site ordered double perovskites $Sr_2CuTeO_6$ and $Sr_2CuWO_6$ are excellent realizations of the $S = 1/2$ square-lattice Heisenberg model.[26,27] These compounds crystallize in the tetragonal space group $I4/m$ with complete rocksalt ordering of the $Cu^{2+}$ and $Te^{6+}/W^{6+}$ cations on the *B'* and *B''*-sites as shown in Figure 1a.[28–30] The magnetic interactions in these materials are highly two-dimensional in the *ab*-plane due to a Jahn-Teller distortion and co-operative orbital ordering of the $3d^9$ $S = 1/2$ $Cu^{2+}$ cations.[30–33] Remarkably, despite being isostructural, $Sr_2CuTeO_6$ and $Sr_2CuWO_6$ have distinctly different magnetic interactions and ground states. $Sr_2CuTeO_6$ has the Néel antiferromagnetic structure (Figure 1b) below $T_N = 29$ K with a propagation vector of **k** = (½, ½, 0).[34] This structure is stabilised by the strong nearest-neighbor $J_1$ interaction with $J_1 = -7.18$ meV and $J_2 = -0.21$ meV.[26] In contrast, $Sr_2CuWO_6$ has the columnar antiferromagnetic structure below $T_N = 24$ K with **k** = (0, ½, ½), see Figure 1c.[35] This structure is stabilised by the strong next-nearest-neighbor $J_2$ interaction with $J_1 = -1.2$ meV and $J_2 = -9.5$ meV.[27] The contrasting magnetic interactions of $Sr_2CuTeO_6$ and $Sr_2CuWO_6$ arise from differences in orbital hybridization of the $4d^{10}$ $Te^{6+}$ and $5d^0$ $W^{6+}$ cations. Strong W 5d – O 2p hybridization in Sr$_2$CuWO$_6$ enables the dominant Cu – O – W – O – Cu superexchange along the diagonal $J_2$.[27,36] Conversely, the dominant $J_1$ interaction in Sr$_2$CuTeO$_6$ is a Cu – O – O – Cu superexchange without significant participation from the Te 4d$^{10}$ states.[26] This d$^{10}$/d$^0$ effect is general for magnetic 3d transition metal double perovskites[37,38] including other Cu$^{2+}$ systems.[39–46]

Given that Sr$_2$CuTeO$_6$ (strong $J_1$) and Sr$_2$CuWO$_6$ (strong $J_2$) are located on the opposite sides of the predicted $J_2/J_1 \approx 0.4$–0.6 spin liquid region in the $S = 1/2$ square-lattice Heisenberg model, the solid solution Sr$_2$CuTe$_{1-x}$W$_x$O$_6$ appeared like a natural system to look for the missing quantum spin liquid.[36] Promisingly, a spin-liquid-like state was first observed in the composition $x$ = 0.5, where muon spectroscopy measurements revealed dynamic magnetism down to 19 mK.[47] A $T$-linear term was observed in the low-temperature specific heat, which is typical of spin liquids. Building on this work, the full Sr$_2$CuTe$_{1-x}$W$_x$O$_6$ magnetic phase diagram has been explored (Figure 1c).[48–51] The Néel ordered state is only observed for $x$ = 0–0.02.[34,51] The spin-liquid-like state is observed in the wide region of x = 0.05–0.6,[47–51] which is incompatible with the very narrow region of stability expected for the $S = 1/2$ square-lattice quantum spin liquid. Finally, columnar magnetic order is found for $x$ = 0.7–1.

The suppression of magnetic order in such as wide region in Sr$_2$CuTe$_{1-x}$W$_x$O$_6$ is likely related to the significant Te/W disorder on the $B''$-site. This results in a special type of bond disorder, where the $J_1$ and $J_2$ interactions are effectively switched on and off depending on the local $B''$-cation in the middle of each Cu$^{2+}$ square.[52,53] The ground state in the spin-liquid-like region has been proposed to be a random-singlet state[23–25,54,55]. The random-singlet state is a disorder-induced spin liquid, where spin singlets of different strengths are formed based on the underlying quenched disorder.[23] Partial spin freezing into either a "spin jam" state[56] or patches of Néel and columnar-type correlated spins[53] have also been proposed for this region.

Here we present an average and local scattering investigation of Sr$_2$CuTe$_{1-x}$W$_x$O$_6$ revealing new insights into this unique magnetic system. Using neutron diffraction, we show that the average crystal structure of $x$ = 0.5 with the spin-liquid-like state is tetragonal at low temperatures retaining an undistorted square of $S = 1/2$ Cu$^{2+}$ cations. Our combined neutron and synchrotron X-ray total scattering experiments show the local structure of $x$ = 0.5 is well described by the average structure. However, we are unable to resolve effects of any potential clustering or ordering of Te and W on the $B''$-site. Our neutron diffraction study of the magnetic order in $x$ = 0.9, 0.8 and 0.7 reveals columnar magnetic order with **k** = (0, ½, ½) as expected. However, an additional reflection belonging to the propagation **k** = (0, ½, 0) is observed for x = 0.8 and 0.7. This suggests that Te-doping in the columnar region disturbs the interlayer magnetic interactions and is responsible for the suppression of magnetic order at x ≈ 0.6. Finally, our polarized neutron study of the compositions $x$ = 0.2 and 0.5 reveal distinctly different local spin correlations related to the short-range correlated states above $T_N$ in the parent phases. This suggests $x$ = 0.2 and 0.5 have different ground states, despite both being in the spin-liquid-like region, and a quantum critical point is expected around x ≈ 0.3.

**EXPERIMENTAL METHODS**

Polycrystalline powder samples of Sr$_2$CuTe$_{1-x}$W$_x$O$_6$ with $0 \leq x \leq 1$ were synthesized using a conventional solid-state synthesis method as described previously.[48] Stoichiometric quantities of SrCO$_3$, CuO, TeO$_2$ and WO$_3$ (≥99.995%, Alfa Aesar) were ground in an agate mortar. The precursor mixture was calcined in air at 900 °C for 12 hours. Synthesis was carried out in air at 1050 °C for 72 h with intermittent grindings.

Time-of-flight neutron total scattering experiments were carried out at the POLARIS diffractometer[57] at the ISIS Neutron and Muon Source. Approximately 11 g of Sr$_2$CuTe$_{0.5}$W$_{0.5}$O$_6$ ($x$ = 0.5) powder was sealed in an 8mm vanadium can. Experiments were carried out in a cryostat at temperatures of 5 K, 100 K and 300 K. The empty sample can and cryostat were also measured for background correction. The data were reduced using standard procedures in Mantid[58] to obtain the Bragg scattering patterns for individual detector banks. Rietveld refinement was carried out using FULLPROF[59] and the structural figures were made with VESTA.[60] Synchrotron X-ray total scattering experiments were carried out at the I15-1 diffractometer at Diamond Light

Source using a wavelength of 0.161669 Å (76 keV). A 2D Perkin Elmer XRD4343CT detector was positioned 20 cm away from capillary to maximize the Q-range for optimal PDF data quality. A dark (without X-rays) detector image was collected to determine the dark current contribution and subsequently subtracted from the data; the detector was kept in a constant read-out state and air-cooled with fans to maintain a constant temperature, which led to negligible changes to the dark current contribution during the experiment.

The pair-distribution functions (PDFs) were obtained using the GUDRUN software package. Neutron data collected at 300 K on POLARIS were corrected for absorption, multiple scattering, and background from the sample environment. X-ray scattering data collected at room temperature on I15-1 were corrected for scattering from the empty capillary, Compton scattering, and incident beam polarisation. The neutron PDF was obtained with a maximum wavevector magnitude $Q_{max}$ = 40 Å$^{-1}$ and the X-ray PDF was obtained with $Q_{max}$ = 25 Å$^{-1}$. The PDFs were expressed as $D(r)$, in the notation of ref. [61].

Constant-wavelength neutron diffraction was measured at the high-intensity D20 diffractometer at the Institut Laue-Langevin. 2g of Sr$_2$CuTe$_{1-x}$W$_x$O$_6$ powders with $x$ = 0.7, 0.8 and 0.9 were enclosed in 6 mm vanadium cans. These compositions are known to magnetically order at $T_N$ = 7 K, 11 K and 15 K, respectively.[48] The data were collected at a wavelength of 2.41 Å at temperatures of 2 K and 30 K. The exact wavelength was determined by refining room-temperature neutron diffraction data against laboratory X-ray data for the $x$ = 0.9 composition. The magnetic Bragg scattering was extracted by subtracting the nuclear scattering observed at 30 K from the 2 K data. The magnetic structure of the $x$ = 0.9 sample was refined using FULLPROF.[59] Potential **k**-vectors in the $I4/m$ space group were considered based on the Brillouin zone database[62] of the Bilbao Crystallographic Server[63–65] and the k-search program included in the FULLPROF Suite.[59] Magnetic phase fractions for $x$ = 0.8 and 0.7 were evaluated by refining the two main magnetic peaks while fixing the ordered moment to be the same in both magnetic phases. The scale factor was fixed by first refining the crystal structure of the corresponding sample. The full width at half maximum (FWHM) was evaluated by fitting a single peak Voigt function.

Diffuse magnetic scattering was investigated at the D7 diffuse scattering spectrometer[66,67] at the Institut Laue-Langevin. 11–19g of sample powder was sealed in aluminum cans with inserts to form an annulus shape. The samples were measured using cold neutrons with a wavelength of 4.8 Å ($E_i$ = 3.55 meV). An orange cryostat was used for temperature control. The samples Sr$_2$CuTeO$_6$ ($x$ = 0) and Sr$_2$CuWO$_6$ ($x$ = 1), which magnetically order at 29 K and 24 K, respectively, were measured at 40 K in the short-range correlated state. Sr$_2$CuTeO$_6$ was also measured at 1.5 K, 60 K and 100 K. The spin-liquid-like $x$ = 0.2 and 0.5 samples[47,48] were measured at 1.5 K. The collected data were reduced using LAMP.[68] The data were corrected for polarizer efficiency with a quartz standard and for detector efficiency with a vanadium standard. The vanadium standard was also used to normalize the data to an absolute intensity scale. The magnetic signal was isolated using xyz polarization analysis, which removes the non-magnetic signal (including background).[69] We have previously presented the raw data with limited analysis for $x$ = 0.2 and 0.5 in ref. [53]. The magnetic diffuse scattering was fitted to an equation described later using the non-linear curve fitting tool in OriginPro. The diffuse scattering was also modelled with a Reverse Monte-Carlo (RMC) approach as implemented in SPINVERT using 8 × 8 × 6 supercells.[70] Each SPINVERT analysis was repeated 10 times in order to reduce statistical noise. The experimental neutron data are available online at refs. [71–73].

**RESULTS AND DISCUSSION**

**Low-temperature crystal structure of $x$ = 0.5**

Sr$_2$CuTeO$_6$ and Sr$_2$CuWO$_6$ crystallize in the $B$-site ordered double perovskite structure with the tetragonal space group $I4/m$.[30,32,34,35] The tetragonal symmetry is retained at low temperatures based on neutron diffraction measurements.[34,35] Therefore, the square-lattice of the $S$ =1/2 Cu$^{2+}$ cations remains undistorted at low temperatures, where the quantum magnetism occurs. The $d^{10}/d^0$ doping does not have a significant effect on the room-temperature crystal structure, as the $Sr_2CuTe_{1-x}W_xO_6$ solid solution retains tetragonal symmetry for the full range $0 < x < 1$.[47,48] However, the low-temperature crystal structure has not been reported for the doped samples.

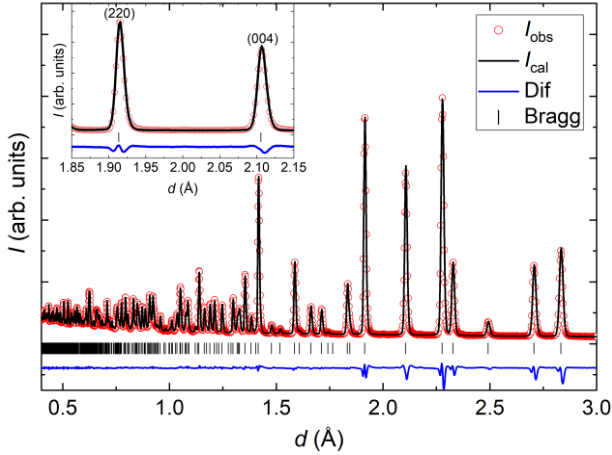

Figure 2. Rietveld refinement of the time-of-flight neutron diffraction data for $Sr_2CuTe_{0.5}W_{0.5}O_6$ ($x$ = 0.5) at 5 K from bank 4 on POLARIS ($2\vartheta$ = 92.59°) with $R_p$ = 2.38% and $R_{wp}$ = 2.29%. The low-temperature structure remains tetragonal in the space group $I4/m$. Inset: Close-up of the (220) and (004) reflections showing that there is no peak splitting or anisotropic line broadening, which confirms the structure is tetragonal.

We investigated the structure of the main spin liquid composition $x$ = 0.5 using neutron diffraction at 5 K, 100 K and 300 K. The space group remains tetragonal $I4/m$ at all temperatures and no structural transition is observed. The refined structure at 300 K (Supporting Information) is essentially identical to the previously published structure based on laboratory X-ray diffraction.[47] Figure 2 shows the refined time-of-flight neutron data for $x$ = 0.5 at 5 K. If the symmetry is lowered from tetragonal, one would expect either the (220) reflection to split or anisotropic broadening of this reflection if the splitting is too small to be resolved. The (220) and (004) reflections are highlighted in the inset. The (220) peak does not split nor is there any anisotropic line broadening. This confirms the average structure remains tetragonal down to 5 K. This means that the $Cu^{2+}$ cations are arranged in a perfect square even at low temperatures. The refined parameters are listed in Table 1. The bond lengths and angles are very similar to the parent phases $Sr_2CuTeO_6$ and $Sr_2CuWO_6$ at low temperatures[34,35] due to the similar ionic radius of $W^{6+}$ and $Te^{6+}$. We do not observe any superlattice reflections arising from Te/W ordering and therefore there is no long-range order of Te and W occupancies on the B''-site.

Table 1. The refined low-temperature crystal structure of $Sr_2CuTe_{0.5}W_{0.5}O_6$ ($x$ = 0.5) at 5 K based on POLARIS time-of-flight neutron data. Space group $I4/m$ with lattice parameters $a$ = 5.41025(10) Å and $c$ = 8.41718(18) Å. $R_p$ = 2.60% and $R_{wp}$ = 2.05% for the high-resolution bank 5 ($2\vartheta$ = 146.72°). The refined crystal structure is shown in Figure 1a with an origin shift of ½ unit cell along $c$.

| Atom | $x$ | $y$ | $Z$ | Occ | $U$ (Å$^2$) |
|---|---|---|---|---|---|
| Sr | 0 | 0.5 | 0.25 | 1 | $U_{11} = U_{22} = 0.0016(2)$, $U_{33} = 0.0017(3)$ |
| Cu | 0 | 0 | 0.5 | 1 | $U_{11} = U_{22} = 0.0004(3)$, $U_{33} = 0.0040(5)$ |
| Te | 0 | 0 | 0 | 0.5 | $U_{11} = U_{22} = 0.0013(4)$, $U_{33} = 0.0012(6)$ |
| W | 0 | 0 | 0 | 0.5 | $U_{11} = U_{22} = 0.0013(4)$, $U_{33} = 0.0012(6)$ |
| O1 | 0.2015(2) | 0.2917(2) | 0 | 1 | $U_{11} = 0.0046(5)$, $U_{22} = 0.0018(5)$, $U_{33} = 0.0047(3)$, $U_{12} = -0.0014(2)$ |
| O2 | 0 | 0 | 0.2267(1) | 1 | $U_{11} = U_{22} = 0.0043(2)$, $U_{33} = 0.0021(4)$ |

**Pair-distribution function analysis of *x* = 0.5**

Our Bragg diffraction data for the *x* = 0.5 composition are well-described by the average structure model with no long-range ordering on the W/Te site. However, this result does not rule out short-range ordering of W and Te. Such short-range ordering would give rise to broad (diffuse) scattering features, which are not modeled in Rietveld refinement. However, such features can be apparent at small distances *r* in the pair-distribution function (PDF), which is the Fourier transform of the diffracted intensity, appropriately normalized and corrected for background scattering.

To investigate this possibility, we analysed our PDF data for *x* = 0.5 using the Topas Academic software[74]. Co-refinements were performed against X-ray and neutron PDF collected at *T* ≈ 300 K. We first performed refinements of the average structure over a large *r*-range ($r_{max}$ = 40 Å) to determine the instrumental parameters $Q_{damp}$ and $Q_{broad}$,[75] which were fixed in subsequent refinements.

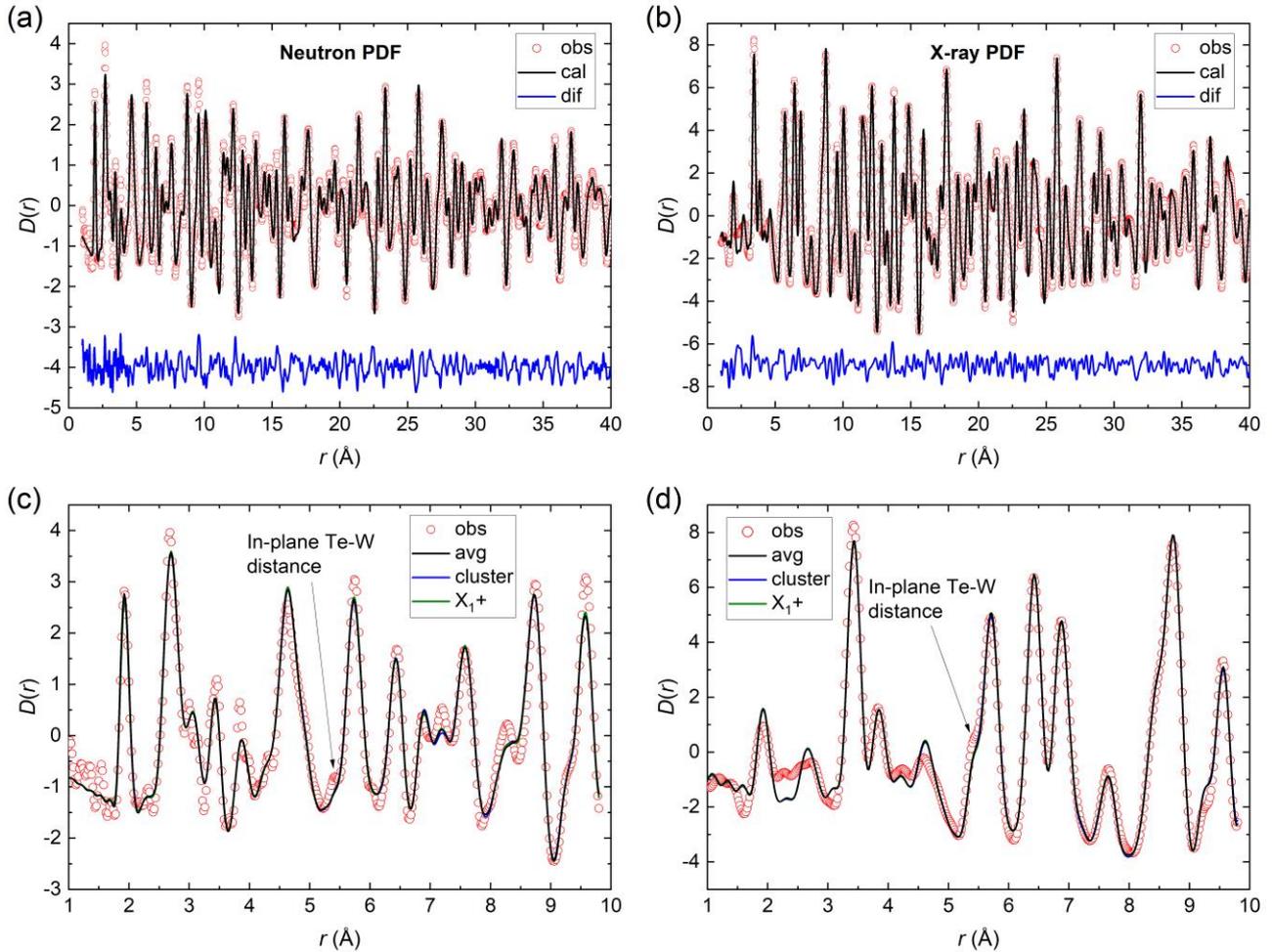

Figure 3. Small-box modelling of the combined neutron and synchrotron X-ray pair-distribution function data for $Sr_2CuTe_{0.5}W_{0.5}O_6$ (*x* = 0.5) at *T* ≈ 300 K. The wide-*r* (1 ≤ *r* ≤ 40 Å) (a) neutron and (b) X-ray PDF data are well described by the average structure. In panels (c) and (d), we fitted the neutron and X-ray PDF in the low-r range (1 ≤ *r* ≤ 9.8 Å) using three different models: a random distribution of Te and W (average structure), clusters of Te and W and finally Te-W ordering ($X_1$+) on the *B''*-sites. Our experiment is not sufficient to differentiate between these models and the fitted lines overlap.

We tested three models of local Te/W occupancies against the neutron+X-ray PDF data: (i) the average-structure model, corresponding to locally random Te/W occupancies; (ii) clustering of Te and W into domains, so that the measured PDF is the average of the PDFs of $Sr_2CuTeO_6$ and $Sr_2CuWO_6$ with identical lattice constants and structural parameters; (iii) a model of local anti-clustering of Te and W, such that W-Te neighbours are favoured over W-W or Te-Te as far as possible. This last model corresponds to the $X_1^+$ irrep for Te/W ordering, and was generated using the ISODISTORT[76] program. For each model, the refined parameters were the scale factors for the two data sets, $a$ and $c$ lattice parameters, O fractional coordinates, isotropic displacement parameters for all atoms, and the low-$r$ peak-sharpening function $d_1$ defined in ref. [75] (13 refined parameters for each model). Refinements were performed for each model over the low-$r$ region (1.0 ≤ $r$ ≤ 9.8 Å), and the wide-$r$ region (1.0 ≤ $r$ ≤ 40 Å).

Our neutron PDF data and fits for the three models described above are shown in Fig. 3c over the 1.0 ≤ $r$ ≤ 9.8 Å range. Good agreement with the data is observed for all models. Unfortunately, however, the goodness-of-fit is not distinguishable between different models, over either the low-$r$ or wide-$r$ regions. That is, the PDF data are equally consistent with correlated $vs.$ random Te/W occupancies. This result is perhaps surprising, since Te and W have reasonable scattering contrast for both neutrons $((b_W/b_{Te})^2 = 0.70)$ and X-rays $((f_W/f_{Te})^2 = 2.03)$, with different contrast ratios for each experiment. However, the situation of physically different models giving rise to essentially identical PDFs is not unknown in the literature.[77] We hypothesise that this situation is more likely in materials where the disordered atoms occupy a high-symmetry site, such as the distorted face-centred cubic lattice of We/Te atoms, where the limitations of powder data are likely to be more important.

Overall, the average-structure model yields good agreement with the measured PDF at low $r$, which suggests that the O1 and Cu atoms do not undergo large "size effect" displacements depending on their local Te/W coordination. In support of this conclusion, we do not observe any anomalously large displacement parameters in our Rietveld refinement, see Table 1. The displacement parameter for O1 is somewhat elongated along $c$, and the Cu displacement parameter is highly cigar-shaped, yet not anomalously large, with $u_{33}(Cu) = 0.0051(5)$ Å$^2$ at 100 K. For comparison, a well-studied quantum-spin-liquid candidate material with occupational disorder, $YbMgGaO_4$, has a much larger $u_{33}(Yb) = 0.0240(4)$ Å$^2$ at 100 K,[78] which implies that local Cu displacements in $Sr_2CuTe_{0.5}W_{0.5}O_6$ are small compared to local Yb displacements in $YbMgGaO_4$. The observation of minimal local distortions in $Sr_2CuTe_{0.5}W_{0.5}O_6$ is consistent with the fact that $Sr_2CuTeO_6$ and $Sr_2CuWO_6$ have nearly identical crystal structures.

**Magnetic order in the W-rich materials $x$ = 0.9, 0.8, 0.7**

One of the main reasons the $Sr_2CuTe_{1-x}W_xO_6$ system is so interesting is the fact that the parent phases $Sr_2CuTeO_6$ ($x$ = 0) and $Sr_2CuWO_6$ ($x$ = 1) have different magnetic structures: $x$ = 0 has the Néel structure with **k** = (½, ½, 0) and $x$ = 1 has the columnar structure with **k** = (0, ½, ½).[34,35] The compositions $x$ = 0.9, 0.8 and 0.7 are also known to magnetically order at $T_N$ = 15 K, 11 K and 7 K based on muon measurements.[48] The magnetic structure for these compositions was proposed to be columnar,[48] which was supported by later neutron diffraction experiments revealing the presence of the (0½½) reflection around $|Q|$ = 0.68 Å$^{-1}$.[51] However, diffraction data was only presented for this one reflection. We have reinvestigated the type of magnetic order in $x$ = 0.9, 0.8 and 0.7 using high-intensity neutron diffraction.

The refined magnetic Bragg scattering for $x$ = 0.9 is shown in Figure 4a. Clear magnetic Bragg peaks are observed at positions corresponding to the propagation vector **k** = (0, ½, ½). The magnetic scattering is almost identical to previous reports on $x$ = 1.[35] The propagation vector has only one irreducible representation $\Gamma_2$, which has three basis vectors along $a$, $b$ and $c$. Similar to $x$ = 1, setting the moment along $a$ resulted in the best fit. The refined $Cu^{2+}$ ordered moment was found to be 0.45(1) $\mu_B$, which is slightly lower than the 0.57(1) $\mu_B$ reported for $x$ = 1.[35] This magnetic structure is shown in the Figure4b inset. A comparison of the magnetic Bragg scattering for $x$ = 0.9, 0.8 and $x$ = 0.7 is presented in panel (b). The main peak at $|Q|$ = 0.68 Å corresponding to (0½½) is still observed in $x$ = 0.8 and 0.7, but the magnetic scattering is significantly weaker and the other reflections of corresponding to **k** = (0, ½, ½) can no longer be resolved. Surprisingly, we observe significant magnetic scattering at $|Q|$ = 0.58 Å$^{-1}$ corresponding to (0½0). This reflection is not allowed for the propagation vector **k** = (0, ½, ½), and therefore this reflection must belong to another propagation vector. Very weak scattering at this position also occurs in the $x$ = 0.9 sample.

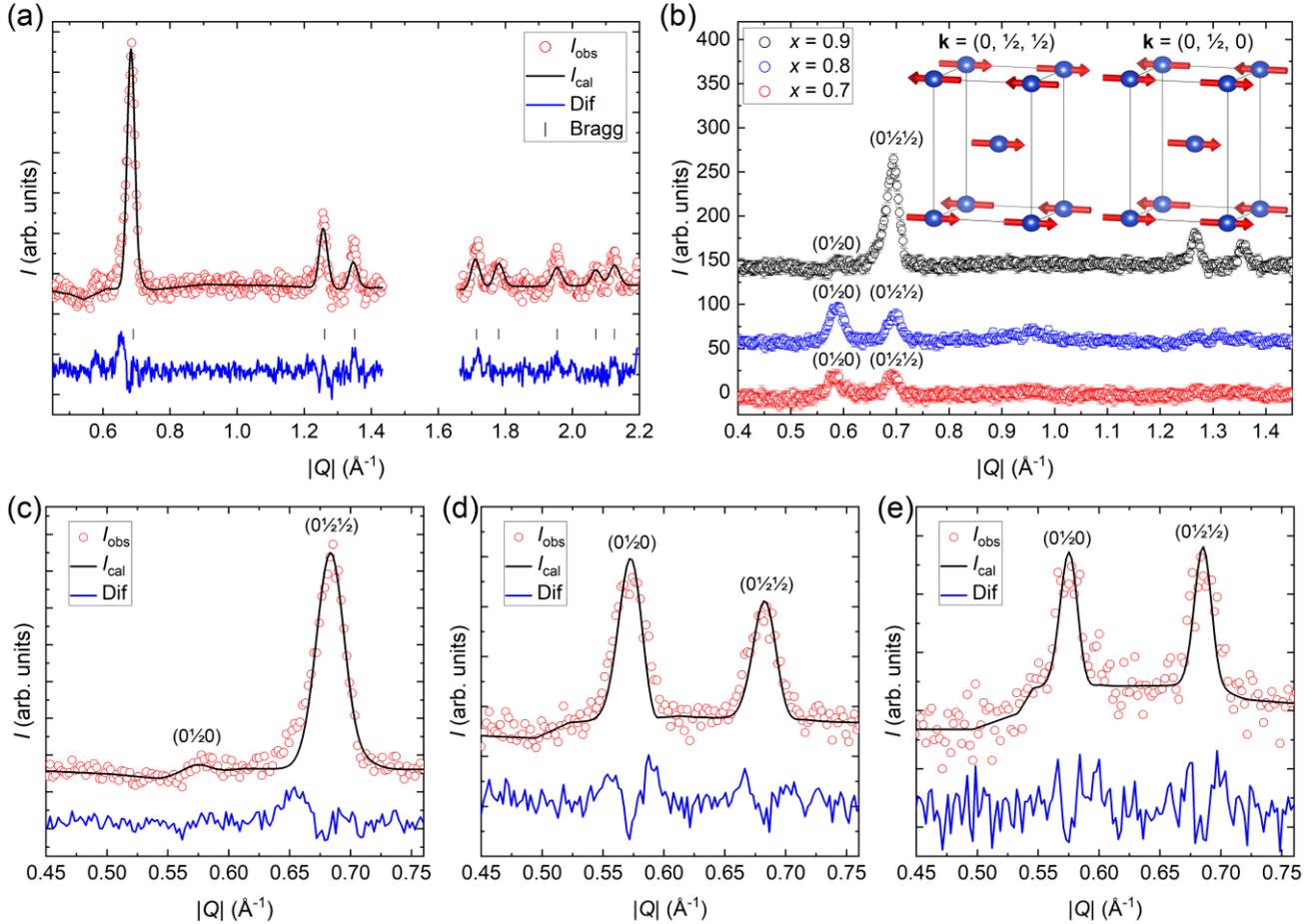

Figure 4. (a) Rietveld refinement of the magnetic neutron diffraction data for Sr$_2$CuTe$_{0.1}$W$_{0.9}$O$_6$ ($x$ = 0.9) at 2 K obtained by subtracting the nuclear scattering measured at 30 K. The propagation vector is **k** = (0, ½, ½) with a refined moment of 0.45(1) $\mu_B$ along $a$ direction ($R_{mag}$ = 13.3%). (b) Comparison of the magnetic neutron diffraction data for Sr$_2$CuTe$_{1-x}$W$_x$O$_6$ samples $x$ = 0.9, 0.8 and 0.7 at 2 K. The magnetic scattering has been normalised using the nuclear scale factor to account for differences in sample masses and counting times. A number of magnetic Bragg peaks are observed for $x$ = 0.9 with the main peak being (0½½). For $x$ = 0.8 and 0.7, intensity of the magnetic reflections is significantly reduced. The (0½½) peak is retained for these compositions, but a new peak at (0½0) is observed. Panel (b) inset: Magnetic structure of $x$ = 1 and 0.9 with propagation vector **k** = (0, ½, ½) and the proposed magnetic structure for the second magnetic phase in x = 0.8 and 0.7 with propagation vector **k** = (0, ½, 0). The ordering along $c$ changes from antiferromagnetic to ferromagnetic. Bottom panels: two-phase magnetic refinement of Sr$_2$CuTe$_{1-x}$W$_x$O$_6$ samples (c) $x$ = 0.9, (d) $x$ = 0.8 and (e) $x$ = 0.7. The $x$ = 0.9 sample is almost entirely in the **k** = (0, ½, ½) magnetic structure that is also observed in $x$ = 1. For $x$ = 0.8 we obtain phase fractions of 65(5)% **k** = (0, ½, 0) and 35(5)% **k** = (0, ½, ½) and for $x$ = 0.7 we find 55(8)% **k** = (0, ½, 0) and 45(8)% **k** = (0, ½, ½).

The magnetic propagation vectors of materials often correspond to high-symmetry points of the Brillouin zone, and therefore these are an excellent starting point for searching for reasonable **k**-vectors. The columnar magnetic structure of $x$ = 1 and 0.9 with **k** = (0, ½, ½) corresponds to the high-symmetry point X. While one might expect **k** = (0, ½, 0) observed in $x$ = 0.8 and 0.7 to also be a high-symmetry point of the Brillouin zone, it is not one in the space group *I4/m*. This is due to the *I*-centering of the lattice and the relationship between the conventional unit cell and the primitive cell. As a consequence, this magnetic structure belongs to the line SM with **k** = (0, *a*, 0) and *a* = 0.500(2).

Symmetry-allowed magnetic structures for **k** = (0, ½, 0) were evaluated using BASIREPS.[59] Two irreducible representations were found: $\Gamma_{mag} = \Gamma_1 + \Gamma_2$. $\Gamma_1$ corresponds to magnetic moments along *c*, while $\Gamma_2$ corresponds to moments within the *ab* plane. Given that we were only able to resolve the main magnetic peak, we are unable to determine the moment directions. Our proposed magnetic structure for x = 0.8 and 0.7 with **k** = (0, ½, 0) is presented in Figure 4b inset with magnetic moments along *a*. This structure corresponds to columnar antiferromagnetic order in the square-layers in the *ab* plane similar to *x* = 1. However, the interlayer coupling along *c* is now ferromagnetic instead of antiferromagnetic.

The presence of the forbidden (0½0) reflection in the magnetic scattering of $Sr_2CuTe_{1-x}W_xO_6$ samples *x* = 0.8 and 0.7 can be interpreted in two ways. It can be due to magnetic phase separation such that parts of the sample have the **k** = (0, ½, ½) and parts have the **k** = (0, ½, 0) magnetic structure. The other possibility is that there is a complex multi-**k** structure that accounts for all the observed magnetic scattering. The weak magnetic scattering makes this distinction complicated, since we can only observe two magnetic Bragg peaks. In the case of magnetic phase separation, we would expect the intensity of the two observed magnetic peaks to vary freely between samples. If the materials have a multi-**k** structure, the relative intensities should remain the same and be an integer ratio.

We investigated this by integrating over the (0½0) and (0½½) reflections. The intensity ratios $A_{(0½0)}/A_{(0½½)}$ were found to be 0.07 for *x* = 0.9, 1.36 for *x* = 0.8 and 1.13 for *x* = 0.7. The intensity ratio of the (0½0) and (0½½) peaks changes with *x*, which suggests magnetic phase separation rather than a multi-**k** structure. In order to evaluate the magnetic phase fractions in the samples, two-phase magnetic refinement of the main magnetic peaks was carried out using FULLPROF, see Figure 4 panels (c) (d) and (e). The *x* = 0.9 sample is almost entirely in the **k** = (0, ½, ½) magnetic structure shared by the parent phase $Sr_2CuWO_6$ (*x* = 1) with only 5(3)% of **k** = (0, ½, 0). For *x* = 0.8 the phase fractions are 35(5)% **k** = (0, ½, ½) and 65(5)% **k** = (0, ½, 0) with an ordered moment of 0.34(1) $\mu_B$. The magnetic scattering in x = 0.7 is weak leading to higher uncertainties of 45(8)% **k** = (0, ½, ½) and 55(8)% **k** = (0, ½, 0) and an ordered moment of only 0.24(1) $\mu_B$.

The widths of the magnetic Bragg peaks are wider than the nuclear peaks in these materials. For *x* = 0.9 we obtain FWHMs of 0.64(2)° for the main magnetic peak (0½½) and 0.46(1)° for the first nuclear peak (011). The FWHMs for *x* = 0.8 are 0.69(3)° and 0.68(5)° for the (0½0) and (0½½) magnetic peaks and 0.47(1)° for (011) and for *x* = 0.7 we obtain 0.51(8)°, 0.55(6)° and 0.47(1)° respectively. The nuclear peak widths are dominated by instrumental broadening as opposed to sample broadening. This supports the presence of size broadening for the magnetic peaks: the size of the magnetic domains is smaller than the crystallite size. Clustering into Te-rich and W-rich regions could lead to the presence of both magnetic structures within a single crystallite. Unfortunately, we were unable to determine the presence or absence of clustering in the X-ray PDF analysis.

The magnetic interactions in $Sr_2CuTe_{1-x}W_xO_6$ materials are highly two-dimensional. The interlayer interaction $J_c$ is an extended Cu – O – W/Te – O – Cu superexchange similar to $J_2$, but much weaker due to the fully occupied Cu $d_{z^2}$ orbitals. In $Sr_2CuWO_6$ (*x* = 1), the interlayer exchange is antiferromagnetic with $J_c$ = -0.01 meV, while the dominant in-plane exchange is $J_2$ = -9.5 meV. As a result of the weak $J_c$, inelastic scattering in the forbidden (0½0) position is observed even in undoped $Sr_2CuWO_6$.[27] The $J_c$ interaction is ferromagnetic in $Sr_2CuTeO_6$ (*x* = 0) following the trend observed in the sign of $J_2$. As a result, it appears that Te-doping (decreasing *x*) disrupts the interlayer interactions leading to the appearance of the competing **k** = (0, ½, 0) structure. Similar magnetic phase separation into structures with different interlayer orderings is also observed in the solid solution $Sr_2Cr_{1.85}Mn_{1.15}As_2O_2$, which has a square-layer of $Cr^{2+}$ (*S* = 2) cations.[79] The weak interlayer interaction is ultimately responsible for magnetic ordering in the $Sr_2CuTe_{1-x}W_xO_6$ materials, since magnetic order in two dimensions at non-zero temperature is forbidden by the Mermin-Wagner theorem when the interactions are isotropic.[80] As such, the disorder in $J_c$ could be the cause of the quantum phase transition from columnar magnetic order in the W-rich side to the spin-liquid-like state at $x \approx 0.6$. This is supported by our diffuse magnetic scattering analysis on $x = 0.5$ and $x = 0.2$ samples (Supporting Information), which show that the average interlayer spin correlations are weak and antiferromagnetic in $x = 0.5$, but change to weakly ferromagnetic for $x = 0.2$.

**Diffuse magnetic scattering in the spin-liquid-like materials $x = 0.2$ and $0.5$**

The $Sr_2CuTe_{1-x}W_xO_6$ parent phases $x = 0$ ($Sr_2CuTeO_6$) and $x = 1$ ($Sr_2CuWO_6$) are known to have short-range correlated magnetic states above $T_N$ based on inelastic neutron scattering studies.[26,27] Spin correlations persist up at least $2T_N \approx 60$ K in both compounds. We previously proposed that the spin-liquid-like state in $x = 0.5$ could be related to the columnar-type short-range correlated state in $x = 1$ based on inelastic neutron scattering data.[52] Similarly, the spin-liquid-like state between $x = 0.05$ and $0.2$ has been proposed to have Néel-type short-range correlations.[51] We can test this hypothesis by measuring the diffuse magnetic scattering in the spin-liquid-like phases and in the short-range correlated states of the parent phases. This allows us to model the local spin correlations in these materials for the first time.

The diffuse magnetic scattering of $Sr_2CuTe_{1-x}W_xO_6$ samples extracted from our D7 experiment is shown in Figure 5. The parent phases $x = 0$ (a) and $x = 1$ (b) were measured above $T_N$ at 40 K, where they are in a short-range correlated magnetic state. The other samples measured were $x = 0.2$ (c) and $x = 0.5$ (d) at 1.5 K, which are in the spin-liquid-like region and lack long-range magnetic order. The diffuse scattering for $x = 0$ at 40 K has a peak around $|Q| \approx 0.85$ Å$^{-1}$. This is related to the Néel magnetic order at low temperatures with the Bragg reflection (½½0) at $|Q| = 0.82$ Å$^{-1}$.[34] The diffuse scattering for $x = 1$ at 40 K is very different, with a main peak around $|Q| \approx 0.6$ Å$^{-1}$. This is related to the columnar magnetic order below $T_N$, which has a magnetic Bragg reflection (0½½) at $|Q| = 0.68$ Å$^{-1}$ and inelastic scattering at the forbidden (0½0) position at $|Q| = 0.58$ Å$^{-1}$.[27] The scattering from the $x = 0.2$ sample at 1.5 K is similar to $x = 0$, although the peak at $|Q| \approx 0.85$ Å$^{-1}$ is significantly broader in $x = 0.2$ with FWHMs of 0.27(8) Å$^{-1}$ and 0.61(15) Å$^{-1}$ respectively. This supports the hypothesis that spin correlations in $x = 0.2$ are Néel-like. The diffuse scattering for $x = 0.5$ is similar to $x = 1$ with a peak around $|Q| \approx 0.6$ Å$^{-1}$. This supports the hypothesis that spin correlations in x = 0.5 are columnar-like.

The low incident energy of $E_i = 3.55$ meV is a limitation in our D7 experiment. Ideally, the incident energy would be high enough to capture all features of the inelastic neutron scattering spectrum.[81] In the case of $Sr_2CuTe_{1-x}W_xO_6$, significant inelastic scattering is observed up to at least 15 meV.[26,27,53] Fitting the high-temperature (100 K) diffuse scattering of $x = 0$ to the $Cu^{2+}$ magnetic form factor results in $\mu_{eff}^2 = 0.91(2)$ $\mu_B^2$, which is only one third of the expected $\mu_{eff}^2 = 3$ $\mu_B^2$ for a $S = 1/2$ cation. We can estimate the effect of the missing magnetic scattering by integrating inelastic neutron scattering data up to higher energies. These cuts for $x = 0.2$ and $0.5$ (Supporting Information) show that our D7 experiment does capture the essential features of the diffuse scattering.

In order to quantify the spin correlations in $Sr_2CuTe_{1-x}W_xO_6$, the observed magnetic scattering cross-sections were fitted to:[82]

$$\left(\frac{d\sigma}{d\Omega}\right)_{mag} = \frac{2}{3}(\gamma_n r_0)^2 \left(\frac{1}{2}F(Q)\right)^2 \mu_{eff}^2 \left(1 + \sum_i Z_i \langle \boldsymbol{S}_0 \cdot \boldsymbol{S}_i \rangle \times \frac{\sin(Qr_i)}{Qr_i}\right) \quad (1)$$

where $F(Q)$ is the magnetic form factor of $Cu^{2+}$, $\mu_{eff}^2 = g^2 S(S+1)$, $Z_i$ is the number of neighboring spins at distance $r_i$ and $\langle \boldsymbol{S}_0 \cdot \boldsymbol{S}_i \rangle$ is the average spin correlation between a central spin and its neighbors at distance $r_i$. The spin correlations have been normalised such that $\langle \boldsymbol{S}_0 \cdot \boldsymbol{S}_i \rangle = 1$ (-1) corresponds to complete (anti-)ferromagnetic alignment of spins. We considered only the nearest-neighbor spins at $r_1 \approx 5.4$ Å and the in-plane next-nearest-neighbor spins at $r_2 \approx 7.6$ Å. These correspond to the spins along the side ($J_1$ interaction) and diagonal ($J_2$ interaction) of the square, respectively. Thus, we have three fitting parameters: $\mu_{eff}^2$ and the spin correlations $\langle S_0 \cdot S_1 \rangle$ and $\langle S_0 \cdot S_2 \rangle$. For complete Néel antiferromagnetic order $\langle S_0 \cdot S_1 \rangle = -1$ and $\langle S_0 \cdot S_2 \rangle = 1$. For columnar order $\langle S_0 \cdot S_1 \rangle = 0$ and $\langle S_0 \cdot S_2 \rangle = -1$.

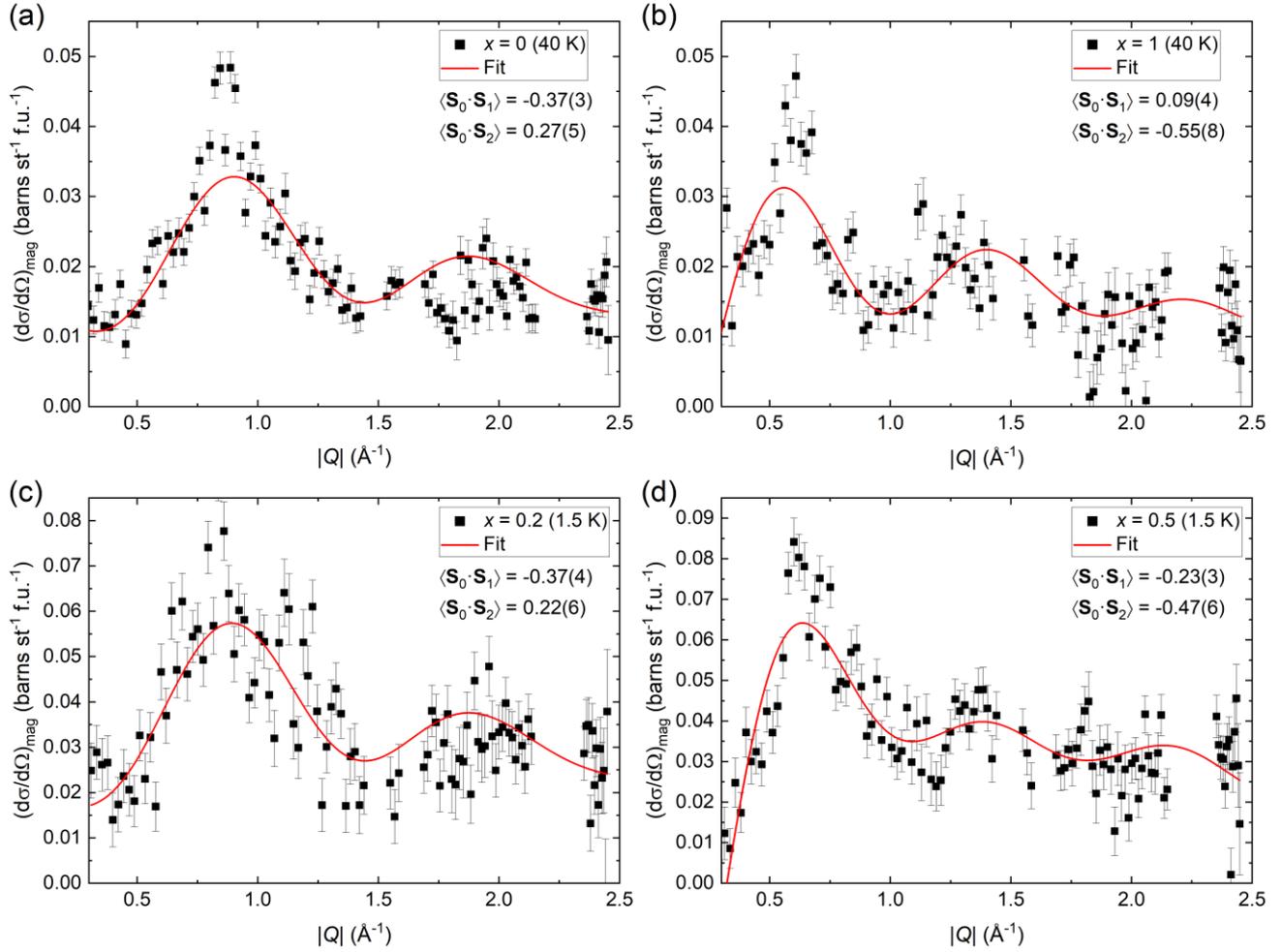

Figure 5. Magnetic diffuse scattering of the $Sr_2CuTe_{1-x}W_xO_6$ parent phases (a) $x = 0$ ($Sr_2CuTeO_6$) and (b) $x = 1$ ($Sr_2CuWO_6$) at 40 K in the short-range correlated state above $T_N$ and the spin-liquid-like phases (c) $x = 0.2$ and (d) $x = 0.5$ at 1.5 K. The red lines are fits to equation (1). The magnetic diffuse scattering has a peak at $|Q| \approx 0.85$ Å$^{-1}$ for $x = 0$ and at $|Q| \approx 0.6$ Å$^{-1}$ for $x = 1$. These are related to the Néel and columnar magnetic ordering below $T_N$, respectively. The $x = 0.2$ data has a peak at $|Q| \approx 0.85$ Å$^{-1}$ similar to $x = 0$ above $T_N$. For $x = 0.5$, the peak is observed at $|Q| \approx 0.6$ Å$^{-1}$ similar to $x = 1$ above $T_N$. The diffuse magnetic scattering in the two spin-liquid-like samples $x = 0.2$ and $x = 0.5$ is clearly different, but also clearly related to the two parent phases above $T_N$.

The fits to equation (1) are shown in Figure 5. The fitting captures the main features of the scattering for all samples, but the model is missing significant intensity at the main peak positions at $|Q| \approx 0.85$ Å$^{-1}$ ($x = 0$) and $|Q| \approx 0.6$ Å$^{-1}$ ($x = 0.5$ and 1). The magnetic diffuse scattering becomes narrower and more Bragg-like when the temperature approaches $T_N$, which is not captured by this model. This could explain why the main peak intensity does not fit well for $x = 0$ and $x = 1$. It should be noted that equation (1) is a simple model including only the two nearest-neighbor in-plane correlations, which are assumed to be fully independent of each other, and Heisenberg spins.[70] Moreover, our experiment does not capture the full spectral weight, which makes the fitting more difficult.

The obtained spin correlations are plotted in Figure 6. For $x = 0$ at 40 K we obtain $\langle S_0 \cdot S_1 \rangle = -0.37(3)$ and $\langle S_0 \cdot S_2 \rangle = 0.27(5)$. These can be characterised as Néel-type correlations linked to the Néel magnetic order in $x = 0$ at low temperatures, where $\langle S_0 \cdot S_1 \rangle = -1$ and $\langle S_0 \cdot S_2 \rangle = 1$. The spin correlations in $x = 0.2$ at 1.5 K are very similar to $x = 0$ at 40 K with $\langle S_0 \cdot S_1 \rangle$ = -0.37(4) and $\langle S_0 \cdot S_2 \rangle$ = 0.22(6) and therefore Néel-like. A significant change in the spin correlations occurs at $x = 0.5$, which is clearly different from $x = 0$ and 0.2. For $x = 0.5$ at 1.5 K we obtain $\langle S_0 \cdot S_1 \rangle$ = -0.23(3) and $\langle S_0 \cdot S_2 \rangle$ = -0.47(6). The correlations are now mainly columnar-like with some Néel-like nearest-neighbor correlations remaining. The spin correlations in $x = 1$ at 40 K are entirely columnar-like with $\langle S_0 \cdot S_1 \rangle$ = 0.09(4) and $\langle S_0 \cdot S_2 \rangle$ = -0.55(8) consistent with the expected $\langle S_0 \cdot S_1 \rangle$ = 0 and $\langle S_0 \cdot S_2 \rangle$ = -1 for complete columnar ordering.

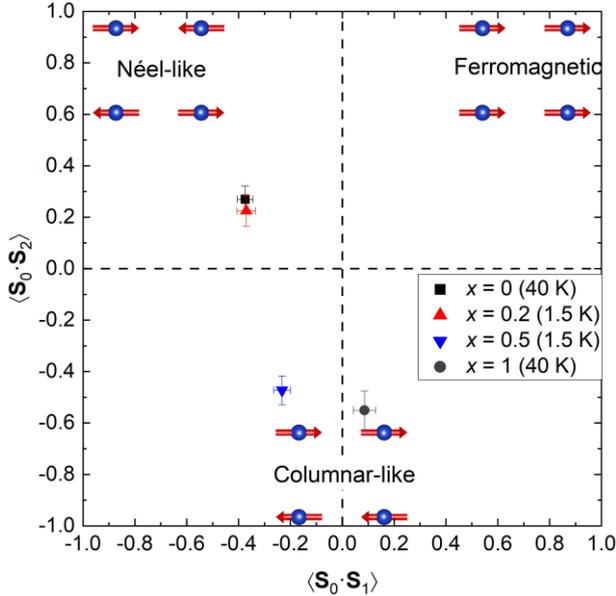

Figure 6. The spin correlations in $Sr_2CuTe_{1-x}W_xO_6$ obtained from fitting the magnetic diffuse scattering to equation (1). The spin correlations in $x = 0$ above $T_N$ are strongly Néel-like with antiferromagnetic $\langle S_0 \cdot S_1 \rangle$ and ferromagnetic $\langle S_0 \cdot S_2 \rangle$. The spin correlations in the spin-liquid-like state in $x = 0.2$ are Néel-like and very similar to $x = 0$. However, the spin-liquid-like state in $x = 0.5$ has very different correlations: weak antiferromagnetic $\langle S_0 \cdot S_1 \rangle$ and strong antiferromagnetic $\langle S_0 \cdot S_2 \rangle$. The latter is a feature of columnar-like spin correlations. Finally, for $x = 1$ above $T_N$ we obtain $\langle S_0 \cdot S_1 \rangle \approx 0$ and strongly antiferromagnetic $\langle S_0 \cdot S_2 \rangle$ as expected of columnar-like spin correlations.

For comparison, we also fitted the diffuse magnetic scattering using a Reverse Monte Carlo (RMC) method as implemented in SPINVERT[70], see Supporting Information. The analysis was complicated by the low quality of the data, which is why we mainly present results from fitting equation (1). The main spin correlations obtained using SPINVERT broadly support our fitting results with the exception of $x = 0.5$, where the weaker $\langle S_0 \cdot S_1 \rangle$ is moderately antiferromagnetic in the equation (1) fit, but zero within experimental error (as expected) in the SPINVERT fit. The same main conclusion of Néel-like correlations in $x = 0.2$ and columnar-like in $x = 0.5$ is observed using both fitting methods.

One advantage of the RMC approach is that spin correlations up to further neighbors can easily be evaluated. The obtained spin correlations for $x = 0.2$ are Néel-type: strong antiferromagnetic nearest-neighbor $\langle S_0 \cdot S_1 \rangle$ correlations ($r_1 \approx 5.4$ Å) and ferromagnetic in-plane next-nearest neighbour correlations $\langle S_0 \cdot S_2 \rangle$ ($r_2 \approx 7.6$ Å). The spin correlations at distance $2r_1$ are weakly ferromagnetic, whereas at $2r_2$ they are nearly zero. The first three in-plane spin correlations are as expected for Néel-type correlations. The interlayer spin correlations are weak and ferromagnetic. For the $x = 0.5$ sample, we obtain $\langle S_0 \cdot S_1 \rangle \approx 0$ and strong antiferromagnetic $\langle S_0 \cdot S_2 \rangle$ correlations. The spin correlations at $2r_1$ and $2r_2$ are ferromagnetic. All four main in-plane spin correlations in $x = 0.5$ are precisely as expected for columnar-type correlations. The interlayer correlations are weak and antiferromagnetic. A comparison of the obtained and expected spin correlations is provided in the Supporting Information.

To summarise, the spin-liquid-like states in $Sr_2CuTe_{1-x}W_xO_6$ with $x = 0.2$ and $x = 0.5$ have short-range spin correlations very similar to the parent phases $x = 0$ and $x = 1$, respectively, above $T_N$. The spin correlations in $x = 0$ and 0.2 are Néel-like while the correlations in $x = 0.5$ and 1 are mainly columnar-like. The cross-over from Néel to columnar correlations is likely to occur around $x \approx 0.3$, because the inelastic neutron scattering data for $x = 0.4$ and 0.5 are very similar.[53] It is unclear whether $x = 0.2$ and $x = 0.5$ have the same ground state given the significant differences in spin correlations. Significant Néel-type correlations are expected in the random-singlet state[55] as we observe for $x = 0.2$. The presence of Néel-type correlations in $x = 0.5$ is less certain. It is clear that the main spin correlations are columnar-type in $x = 0.5$, but some weak Néel-like correlations were found in the fits to equation (1). However, we did not observe these correlations in the SPINVERT analysis of the same data. As such, our experiment suggests $x = 0.2$ and $x = 0.5$ have a different ground state and that there is quantum critical point between them. This could be a quantum critical point between two different types of random-singlet states or a random-singlet state and a state with weakly frozen moments.[56] We expect this quantum critical point, should it exist, to occur around $x \approx 0.3$, where the spin correlations change from Néel to columnar. This could be further investigated by muon spectroscopy as the scaling behaviour of the muon spin relaxation rate should change in the presence of a critical point.[51]

**Conclusions**

We have investigated different compositions of the $S = 1/2$ square-lattice antiferromagnet $Sr_2CuTe_{1-x}W_xO_6$ using neutron and X-ray scattering. The average crystal structure of the spin-liquid-like $x = 0.5$ sample remains tetragonal down to 5 K confirming a square magnetic lattice. Our PDF analysis showed the local structure is overall well described by the average structure, although we were unable to distinguish between different models of Te and W correlations (random, clustering or ordering). Our neutron diffraction experiments of the W-rich magnetically ordered materials reveal the presence of columnar magnetic order for $x = 0.7-1$. Surprisingly, magnetic phase separation was observed for $x = 0.7$ and 0.8 with part of the sample ordering ferromagnetically and part antiferromagnetically along $c$, while the columnar order in the $ab$-plane was preserved. This shows replacing W with Te leads to disorder in the interlayer interactions, which could be the origin of the transition to the spin-liquid-like state at $x \approx 0.6$.

The spin correlations of the spin-liquid-like phases $x = 0.2$ and 0.5 were investigated using polarized neutrons. The spin-correlations in $x = 0.2$ are Néel-like, and very similar to the short-range correlated state observed above $T_N$ in $Sr_2CuTeO_6$. The spin correlations in $x = 0.5$ are mainly columnar-type with potentially some weak Néel-like correlations remaining, and similar to the short-range correlated state above $T_N$ in $Sr_2CuWO_6$. Despite both compositions being in the spin-liquid-like region, the spin correlations are very different. This suggests the ground states are also different, because Néel-type correlations are expected in the random-singlet state. If so, a quantum critical point would be expected around $x \approx 0.3$, where the spin correlations change from Néel to columnar-like.

Our results highlight the complexity of $Sr_2CuTe_{1-x}W_xO_6$ as a $S = 1/2$ square-lattice antiferromagnet with frustration and disorder in the local in-plane and interlayer interactions. The strong suppression of Néel order at $x \approx 0.05$ and the related quantum critical point has previously garnered significant attention.[50,51,55] Our results show that further investigation is warranted at higher doping levels both at the other known quantum critical point at x ≈ 0.6 and our proposed new quantum critical point near $x \approx 0.3$. Given that the $d^{10}/d^0$ substitution approach for tuning magnetism is applicable to many 3d transition metal oxides,[37] our insights have relevance to other systems such as $SrLaSb_{1-x}Nb_xCuO_6$ or the spin ladder $Ba_2CuTe_{1-x}W_xO_6$.[40,46]

**Acknowledgements**

The authors thank Dr Lucy Clark, Dr Ross Stewart and Dr Jennifer Graham for fruitful discussions. Dr Clemens Ritter is thanked for technical assistance with the D20 data. O.M. is grateful for funding through Leverhulme Trust Early Career Fellowship ECF-2021-170. O.M. and E.C. acknowledge support from the Leverhulme Trust Research Project Grant RPG-2017-109. E.F. and H.R. acknowledge funding from the European Research Council through the Synergy network HERO (Grant No. 810451). The work of J.A.M.P. was supported by the U.S. Department of Energy, Office of Science, Basic Energy Sciences, Materials Sciences and Engineering Division. The authors thank the Science and Technology Facilities Council for beamtime allocated at ISIS (proposal number RB1810107). We acknowledge Diamond Light Source for access to beamline I15-1 XPDF within the joint (i15-1-POLARIS) PDF scheme under proposal CY21802. The authors are grateful for beamtime on the D20 high-flux diffractometer and the D7 diffuse scattering spectrometer at the Institut Laue-Langevin under proposals 5-31-2638 and 5-32-895.

This manuscript has been coauthored by UT-Battelle, LLC under Contract No. DE-AC05-00OR22725 with the U.S. Department of Energy. The United States Government retains and the publisher, by accepting the article for publication, acknowledges that the United States Government retains a non-exclusive, paid-up, irrevocable, world-wide license to publish or reproduce the published form of this manuscript, or allow others to do so, for United States Government purposes. The Department of Energy will provide public access to these results of federally sponsored research in accordance with the DOE Public Access Plan (http://energy.gov/downloads/doe-public-access-plan).

**Authors contributions**

O.M., P.B., M. K, H.R. and H.W. conceived and planned the study. The samples were synthesised by O.M. and S.V. The POLARIS experiment was carried out by O.M., J.P., H.P. and H.W. with data analysis by J.P. The I15-1 experiment was performed by M. D.-L. and analysed by J.P. The D20 experiment was performed by O.M., E.F., T.H. and H.W. and analysed by O.M. and E.C. The D7 experiments were carried out by O.M., E.F., L M.-T. and H.W. with analysis by O.M. and E.F. Funding for this work was acquired by O.M., E.C. and H.R. The manuscript was written by O.M. with contributions from all authors.

Supporting information

# Structure, spin correlations and magnetism of the $S$ = 1/2 square-lattice antiferromagnet $Sr_2CuTe_{1-x}W_xO_6$ (0 ≤ $x$ ≤ 1)

Otto Mustonen,[1,2]* Ellen Fogh,[3]* Joseph A. M. Paddison,[4] Lucile Mangin-Thro,[5] Thomas Hansen,[5] Helen Playford,[6] Maria Diaz-Lopez,[7] Peter Babkevich,[3] Sami Vasala,[8] Maarit Karppinen,[9] Edmund J. Cussen,[2] Henrik M. Rønnow,[3] Helen C. Walker[6]*

1 School of Chemistry, University of Birmingham, Birmingham B15 2TT, United Kingdom
2 Department of Material Science and Engineering, University of Sheffield, Sheffield S1 3JD, United Kingdom
3 Laboratory for Quantum Magnetism, Institute of Physics, École Polytechnique Fédérale de Lausanne (EPFL), CH-1015 Lausanne, Switzerland
4 Materials Science and Technology Division, Oak Ridge National Laboratory, Oak Ridge, Tennessee 37831, USA
5 Institut Laue Langevin, 71 Avenue des Martyrs, CS 20156, F-38042 Grenoble Cedex 9, France
6 ISIS Neutron and Muon Source, Rutherford Appleton Laboratory, Chilton, Didcot OX11 0QX, United Kingdom
7 Université Grenoble Alpes, CNRS, Grenoble INP, Institut Néel, 38000 Grenoble, France
8 ESRF - The European Synchrotron, 38000 Grenoble, France
9 Department of Chemistry and Materials Science, Aalto University, FI-00076, Espoo, Finland

## Table of contents

# Crystal structure of $Sr_2CuTe_{0.5}W_{0.5}O_6$ ($x = 0.5$)

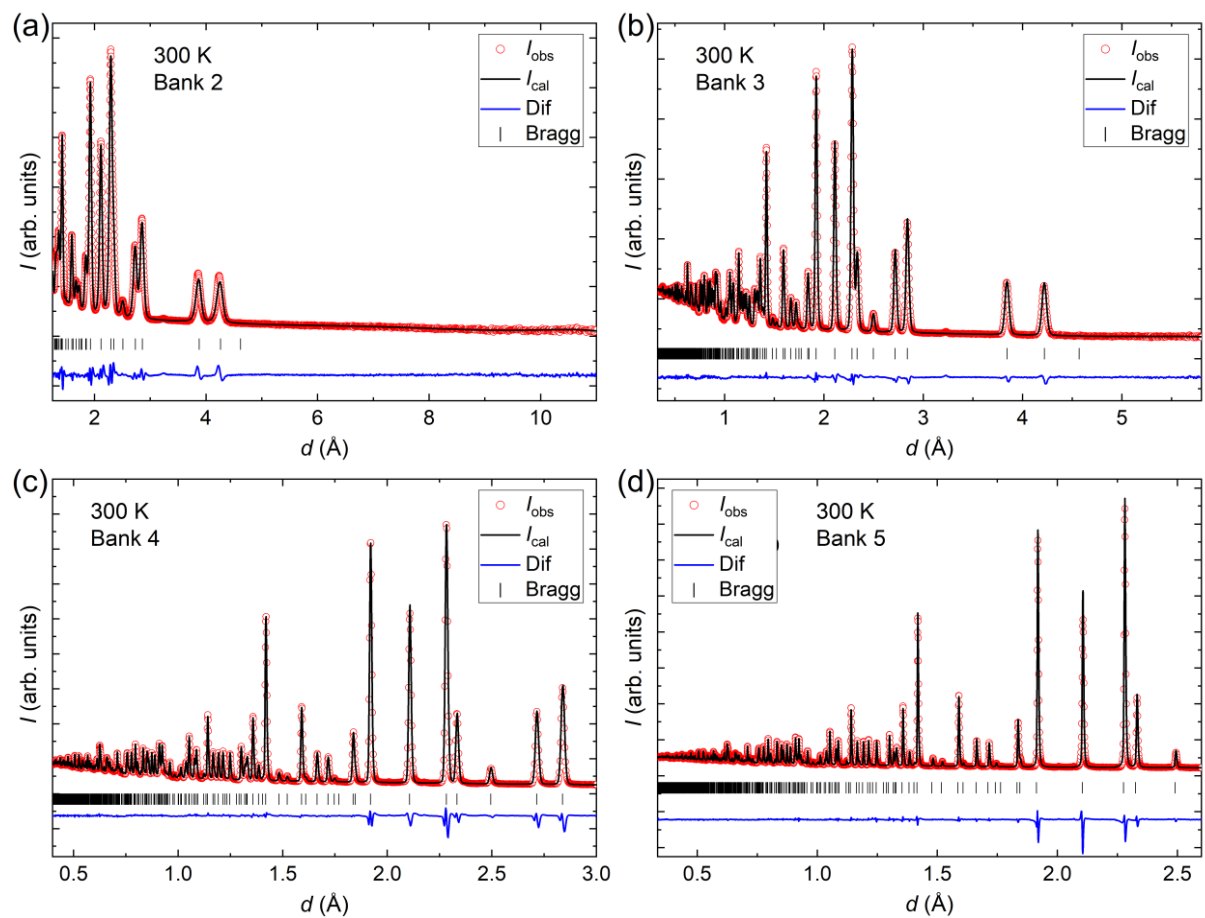

Figure S1. Rietveld refinement of the time-of-flight neutron diffraction data for $Sr_2CuTe_{0.5}W_{0.5}O_6$ ($x = 0.5$) at 300 K collected on POLARIS. **a** Data for bank 2 at $2\vartheta = 25.99°$ with $R_p = 1.55\%$ and $R_{wp} = 2.41\%$. **b** Data for bank 3 at $2\vartheta = 52.21°$ with $R_p = 1.26\%$ and $R_{wp} = 1.54\%$. **c** Data for bank 4 at $2\vartheta = 92.59°$ with $R_p = 2.23\%$ and $R_{wp} = 2.22\%$. **d** Data for bank 5 at $2\vartheta = 146.72°$ with $R_p = 2.19\%$ and $R_{wp} = 1.75\%$.

Table S1. The refined crystal structure of $Sr_2CuTe_{0.5}W_{0.5}O_6$ ($x = 0.5$) at 300 K based on POLARIS time-of-flight neutron data. Space group $I4/m$ with lattice parameters $a = 5.42619(12)$ Å and $c = 8.4239(2)$ Å. $R_p = 2.19\%$ and $R_{wp} = 1.75\%$ for the high-resolution bank 5 ($2\vartheta = 146.72°$).

| Atom | $x$ | $y$ | $z$ | Occ | $U$ (Å$^2$) |
|---|---|---|---|---|---|
| Sr | 0 | 0.5 | 0.25 | 1 | $U_{11} = U_{22} = 0.0061(2)$, $U_{33} = 0.0077(4)$ |
| Cu | 0 | 0 | 0.5 | 1 | $U_{11} = U_{22} = 0.0026(4)$, $U_{33} = 0.0095(6)$ |
| Te | 0 | 0 | 0 | 0.5 | $U_{11} = U_{22} = 0.0035(6)$, $U_{33} = 0.0029(8)$ |
| W | 0 | 0 | 0 | 0.5 | $U_{11} = U_{22} = 0.0035(6)$, $U_{33} = 0.0029(8)$ |
| O1 | 0.2051(3) | 0.2890(2) | 0 | 1 | $U_{11} = 0.0074(7)$, $U_{22} = 0.0040(7)$, $U_{33} = 0.0133(4)$, $U_{12} = -0.0032(2)$ |
| O2 | 0 | 0 | 0.22604(14) | 1 | $U_{11} = U_{22} = 0.0104(3)$, $U_{33} = 0.0046(5)$ |

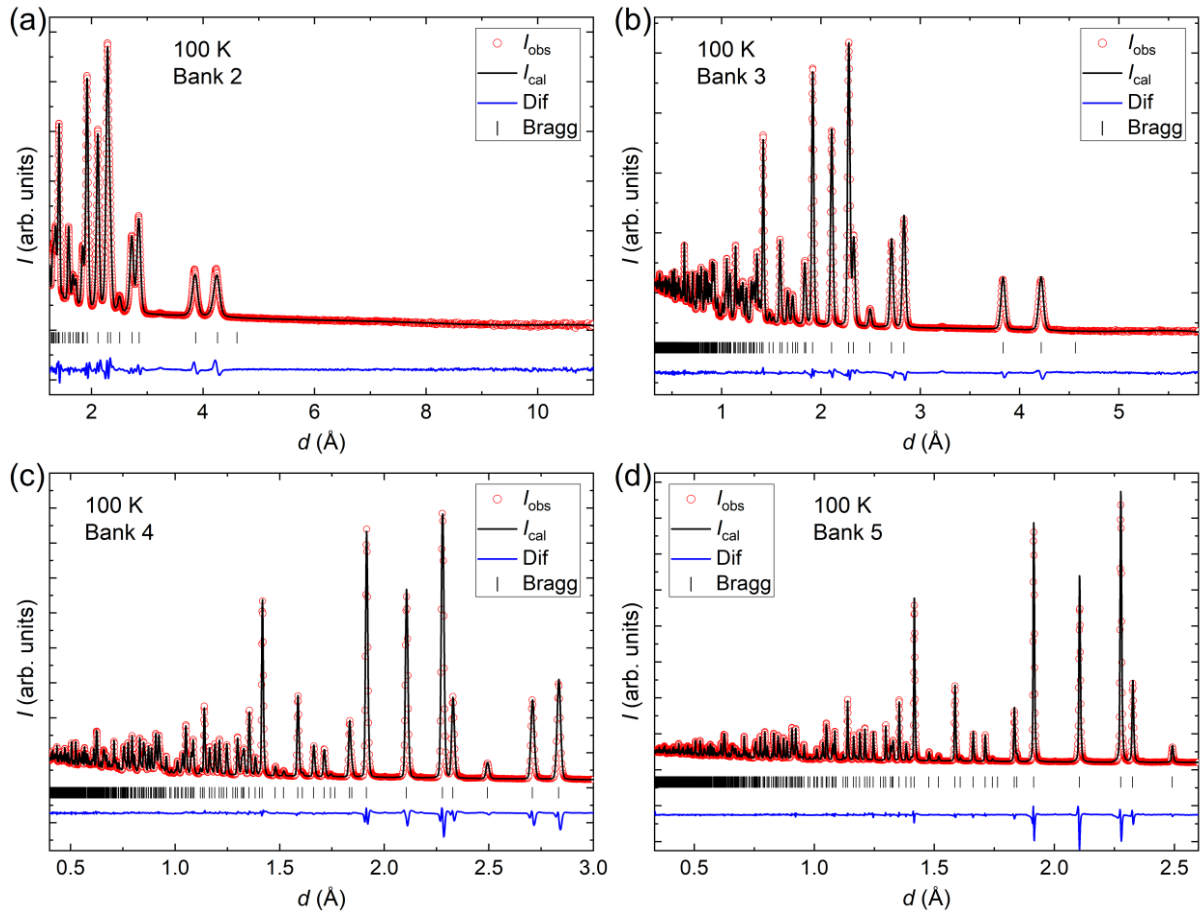

Figure S2. Rietveld refinement of the time-of-flight neutron diffraction data for $Sr_2CuTe_{0.5}W_{0.5}O_6$ (x = 0.5) at 100 K collected on POLARIS. **a** Data for bank 2 at $2\vartheta$ = 25.99° with $R_p$ = 1.51% and $R_{wp}$ = 2.33%. **b** Data for bank 3 at $2\vartheta$ = 52.21° with $R_p$ = 1.25% and $R_{wp}$ = 1.48%. **c** Data for bank 4 at $2\vartheta$ = 92.59° with $R_p$ = 2.36% and $R_{wp}$ = 2.29%. **d** Data for bank 5 at $2\vartheta$ = 146.72° with $R_p$ = 2.53% and $R_{wp}$ = 2.01%.

Table S2. The refined crystal structure of $Sr_2CuTe_{0.5}W_{0.5}O_6$ (x = 0.5) at 100 K based on POLARIS time-of-flight neutron data. Space group $I4/m$ with lattice parameters $a$ = 5.41226(10) Å and $c$ = 8.41815(18) Å. $R_p$ = 2.53% and $R_{wp}$ = 2.01% for the high-resolution bank 5 ($2\vartheta$ = 146.72°).

| Atom | x | y | z | Occ | U (Å$^2$) |
|---|---|---|---|---|---|
| Sr | 0 | 0.5 | 0.25 | 1 | $U_{11}$ = $U_{22}$ = 0.0025(2), $U_{33}$ = 0.0029(3) |
| Cu | 0 | 0 | 0.5 | 1 | $U_{11}$ = $U_{22}$ = 0.0007(3), $U_{33}$ = 0.0051(5) |
| Te | 0 | 0 | 0 | 0.5 | $U_{11}$ = $U_{22}$ = 0.0018(4), $U_{33}$ = 0.0015(7) |
| W | 0 | 0 | 0 | 0.5 | $U_{11}$ = $U_{22}$ = 0.0018(4), $U_{33}$ = 0.0015(7) |
| O1 | 0.2020(2) | 0.2915(2) | 0 | 1 | $U_{11}$ = 0.0050(5), $U_{22}$ = 0.0021(5), $U_{33}$ = 0.0060(2), $U_{12}$ = -0.0016(2) |
| O2 | 0 | 0 | 0.2266(1) | 1 | $U_{11}$ = $U_{22}$ = 0.0052(2), $U_{33}$ = 0.0027(4) |

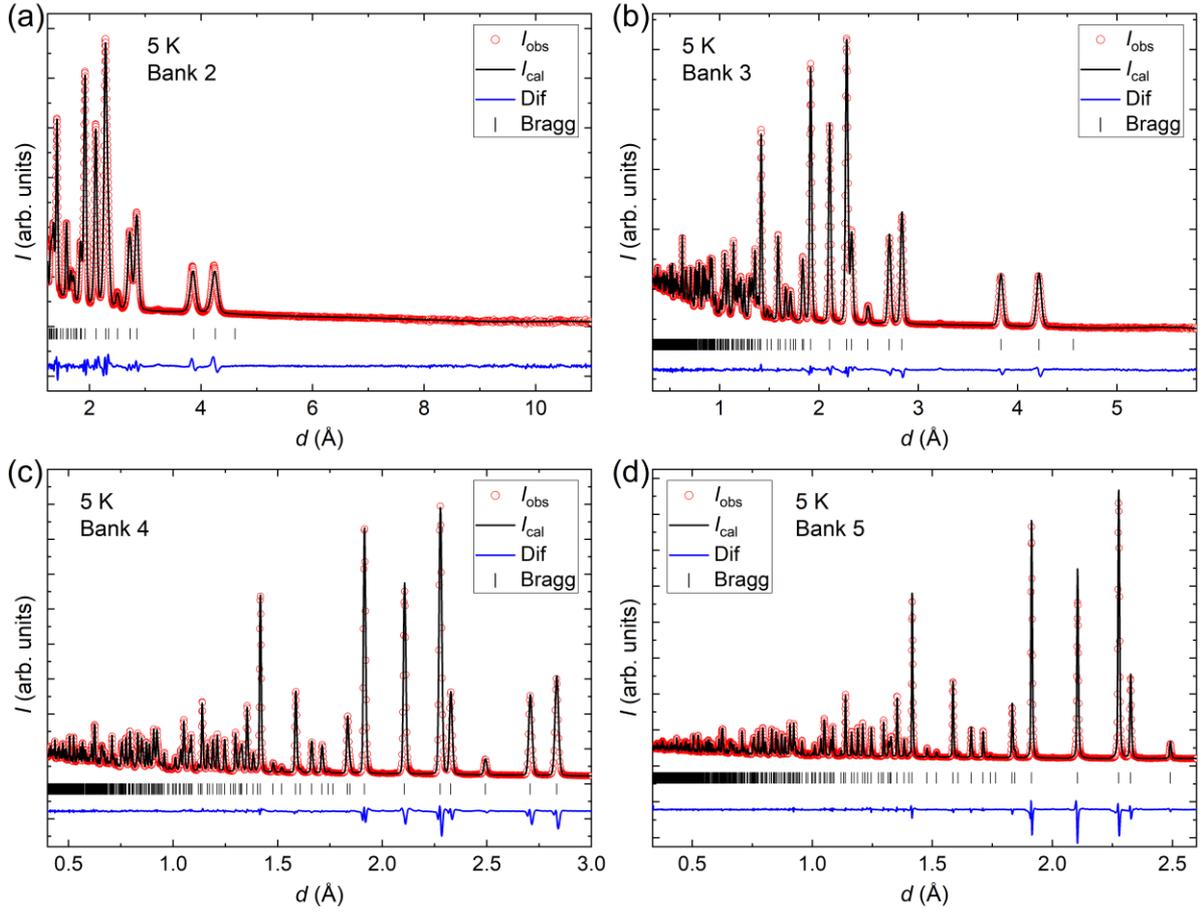

Figure S3. Rietveld refinement of the time-of-flight neutron diffraction data for $Sr_2CuTe_{0.5}W_{0.5}O_6$ ($x$ = 0.5) at 5 K collected on POLARIS. **a** Data for bank 2 at $2\vartheta$ = 25.99° with $R_p$ = 1.72% and $R_{wp}$ = 2.39%. **b** Data for bank 3 at $2\vartheta$ = 52.21° with $R_p$ = 1.24% and $R_{wp}$ = 1.46%. **c** Data for bank 4 at $2\vartheta$ = 92.59° with $R_p$ = 2.38% and $R_{wp}$ = 2.29%. **d** Data for bank 5 at $2\vartheta$ = 146.72° with $R_p$ = 2.60% and $R_{wp}$ = 2.05%.

Table S3. The refined low-temperature crystal structure of $Sr_2CuTe_{0.5}W_{0.5}O_6$ ($x$ = 0.5) at 5 K based on POLARIS time-of-flight neutron data. Space group $I4/m$ with lattice parameters $a$ = 5.41025(10) Å and $c$ = 8.41718(18) Å. $R_p$ = 2.60% and $R_{wp}$ = 2.05% for the high-resolution bank 5 ($2\vartheta$ = 146.72°). This table is reproduced here for convenience from Table 1 in the article.

| Atom | $x$ | $y$ | $z$ | Occ | $U$ (Å$^2$) |
|---|---|---|---|---|---|
| Sr | 0 | 0.5 | 0.25 | 1 | $U_{11} = U_{22}$ = 0.0016(2), $U_{33}$ = 0.0017(3) |
| Cu | 0 | 0 | 0.5 | 1 | $U_{11} = U_{22}$ = 0.0004(3), $U_{33}$ = 0.0040(5) |
| Te | 0 | 0 | 0 | 0.5 | $U_{11} = U_{22}$ = 0.0013(4), $U_{33}$ = 0.0012(6) |
| W | 0 | 0 | 0 | 0.5 | $U_{11} = U_{22}$ = 0.0013(4), $U_{33}$ = 0.0012(6) |
| O1 | 0.2015(2) | 0.2917(2) | 0 | 1 | $U_{11}$ = 0.0046(5), $U_{22}$ = 0.0018(5), $U_{33}$ = 0.0047(3), $U_{12}$ = -0.0014(2) |
| O2 | 0 | 0 | 0.2267(1) | 1 | $U_{11} = U_{22}$ = 0.0043(2), $U_{33}$ = 0.0021(4) |

## Magnetic scattering in the ordered W-rich phases

Table S4. Integrated magnetic Bragg peak areas for the magnetically ordered $Sr_2CuTe_{1-x}W_xO_6$ samples $x$ = 0.9, 0.8 and 0.7. D20 data is shown in main article Figure 5a. The peak area ratio of the (0½0) and (0½½) reflections change with $x$. This suggests the origin of the additional (½00) reflection is magnetic phase separation with two different propagation vectors **k** = (0, ½, 0) and **k** = (0, ½, ½). If the (0½0) reflection arose as part of a more complicated multi-**k** magnetic structure, the (0½0)/(0½½) ratio would be expected to remain constant.

| Sample | (0½0) area | (0½½) area | (0½0)/(0½½) |
|---|---|---|---|
| $x$ = 0.9 | 5.45 | 81.56 | 0.07 |
| $x$ = 0.8 | 20.33 | 15.00 | 1.36 |
| $x$ = 0.7 | 14.69 | 12.96 | 1.13 |

## Magnetic scattering of $Sr_2CuTeO_6$ at various temperatures

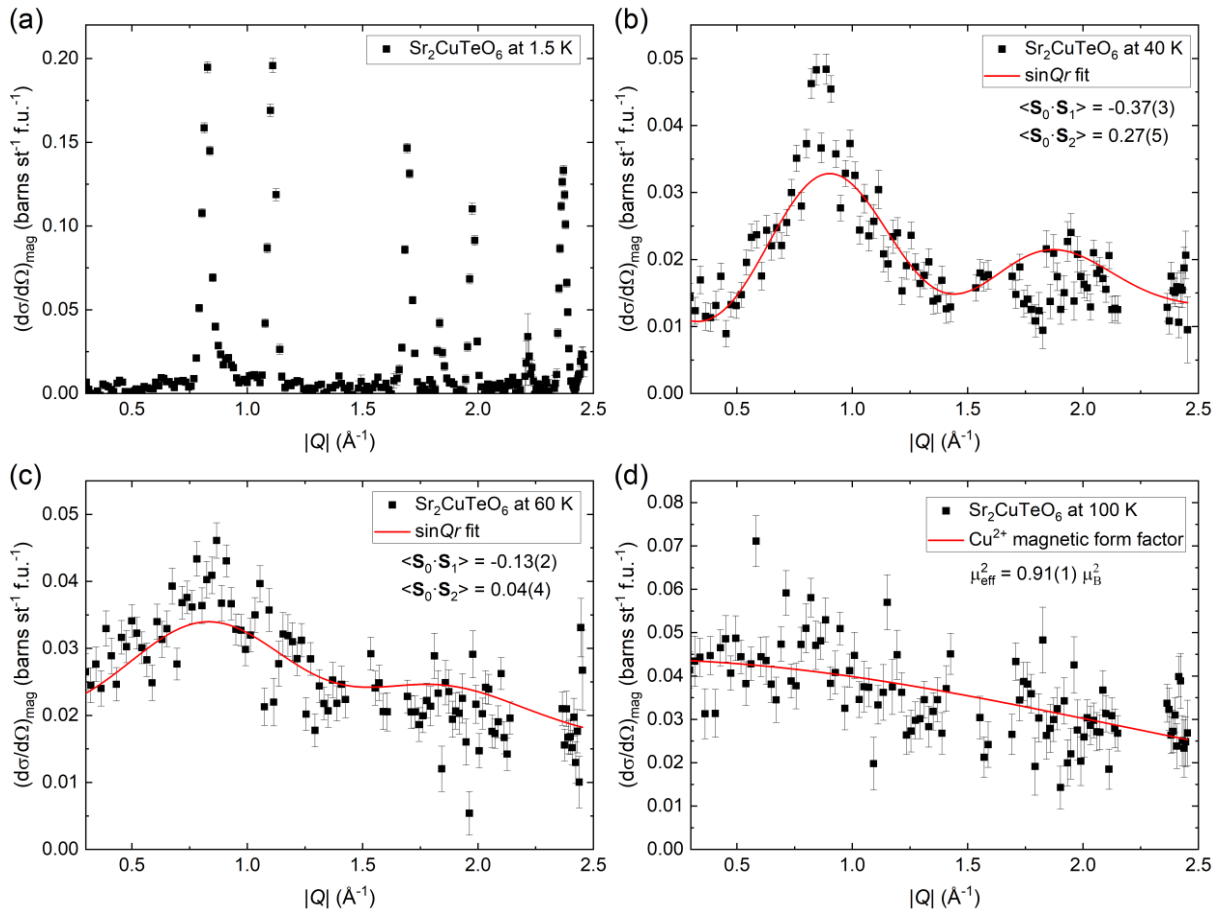

Figure S4. Magnetic scattering of $Sr_2CuTeO_6$ ($x$ = 0) at **a** 1.5 K, **b** 40 K, **c** 60 K and **d** 100 K. Magnetic Bragg peaks are clearly visible in the 1.5 K data as expected below $T_N \approx 29$ K. Diffuse magnetic scattering indicative of Néel-type correlations is observed at 40 K and 60 K. At 100 K, the scattering is mainly paramagnetic, but minor correlations appear to be still present around $|Q|$ = 0.85 Å$^{-1}$. A fit to the Cu$^{2+}$ magnetic form factor yields $\mu_{eff}^2$ = 0.91(1) $\mu_B^2$, which is a third of the expected value. Therefore, our experiment captures only the lowest 1/3$^{rd}$ energy excitations in these compounds.

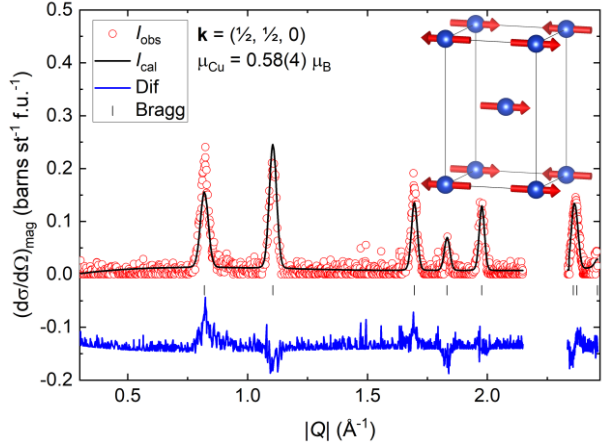

Figure S5. Refined magnetic structure of $Sr_2CuTeO_6$ at 1.5 K. Propagation vector **k** = (½, ½, 0), a moment of 0.58(4) $\mu_B$ per copper and $R_{mag}$ = 23.5 %. The refined moment is slightly smaller than the 0.69(6) $\mu_B$ previously reported in ref. [1], but consistent with the 0.57(1) $\mu_B$ reported for $Sr_2CuWO_6$.[2] Inset: The Néel antiferromagnetic structure of $Sr_2CuTeO_6$. The nearest-neighbor spins order antiferromagnetically along the side of the square, while the in-plane next-nearest-neighbor (square diagonal) and interplane ordering along *c* are ferromagnetic.

## Integrated inelastic neutron scattering data

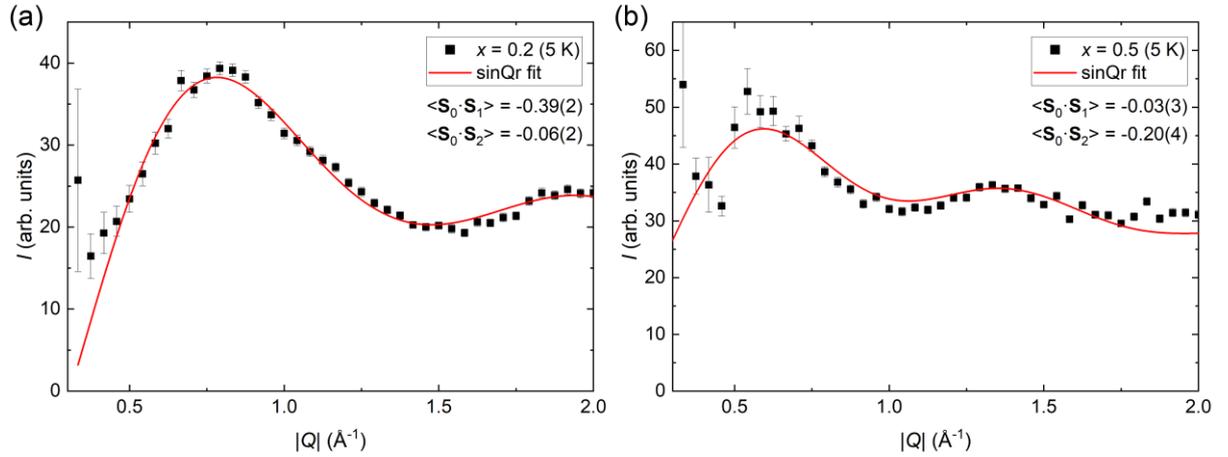

Figure S6. Integrated inelastic neutron scattering between 2.5 meV and 7.5 meV energy transfer for $Sr_2CuTe_{1-x}W_xO_6$ samples **a** $x$ = 0.2 and **b** $x$ = 0.5. Data from refs. [3,4] collected on MERLIN at $T$ = 5 K with $E_i$ = 18 meV. The positions of the maximums and the overall shapes of the curves are very similar to the D7 measurements with $E_i$ = 3.55 meV. This shows that despite not capturing the full spectral weight, the D7 data is representative of the overall scattering at least up to 7.5 meV. The limited low-|$Q$| coverage of the MERLIN data at higher energies prohibits comparisons up to 15-20 meV, where features are still observed in the spectra.

# SPINVERT fits of diffuse magnetic scattering

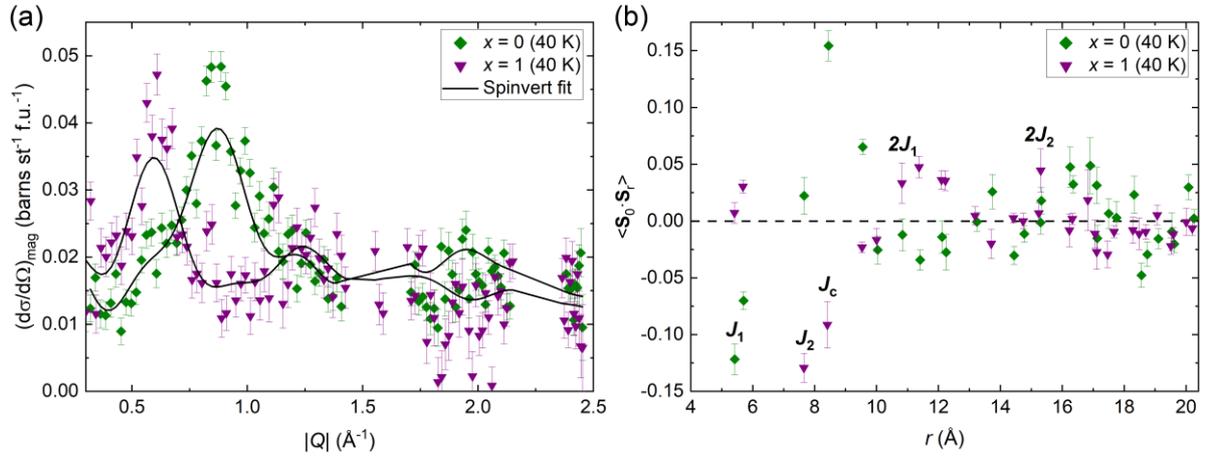

Figure S7. **a** SPINVERT fits of the diffuse magnetic scattering of $Sr_2CuTeO_6$ ($x = 0$) and $Sr_2CuWO_6$ ($x = 1$) at 40 K in the short-range correlated state above $T_N$. The quality of the fits is limited by the very weak magnetic scattering and limited energy coverage. The magnetic scattering of $x = 0$ has a peak at $|Q| \approx 0.85$ Å$^{-1}$, which is related to the (½½0) reflection of the Néel magnetic structure below $T_N$. For $x = 1$, a peak is observed at $|Q| \approx 0.6$ Å$^{-1}$, which is related to the main (0½½) magnetic Bragg peak of the columnar structure and inelastic scattering from the forbidden (0½0) reflection due to the weak interlayer coupling. **b** Spin correlation functions of $Sr_2CuTeO_6$ and $Sr_2CuWO_6$ obtained from SPINVERT fits. The spin correlations for $x = 0$ are Néel-type with antiferromagnetic $J_1$ correlations and ferromagnetic $J_2$ correlations. The correlations in $x = 1$ are columnar-type with very weak $J_1$ and antiferromagnetic $J_2$ correlations.

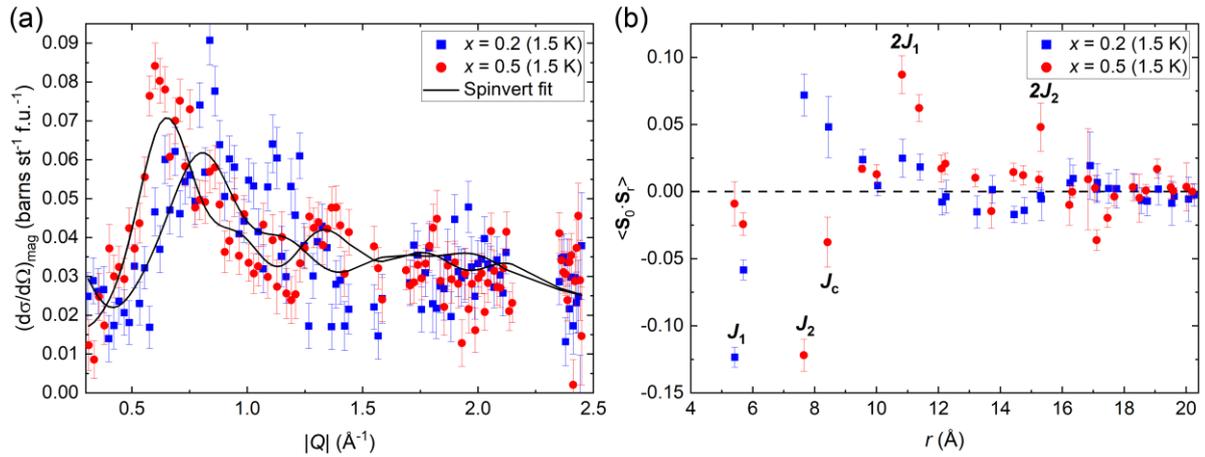

Figure S8. **a** SPINVERT fits to diffuse magnetic scattering of $x = 0.2$ and $0.5$ samples at 1.5 K. The quality of the fits is limited by the very weak magnetic scattering and limited energy coverage. The maximum in scattering is observed at different $|Q|$ positions in the two samples: at $|Q| \approx 0.85$ in $x = 0.2$ and at $|Q| \approx 0.6$ Å$^{-1}$ in $x = 0.5$. This indicates a significant difference in the spin correlations of these two materials. There is a feature in the $x = 0.2$ data at $|Q| \approx 1.13$ Å$^{-1}$ that is not captured in the model. **b** Spin correlation functions of the $x = 0.2$ and $0.5$ samples obtained from SPINVERT fits. Spin correlations in the $x = 0.2$ are Néel-type with strong antiferromagnetic $J_1$ and ferromagnetic $J_2$. The correlations are columnar-type in $x = 0.5$ with negligible $J_1$ and strong antiferromagnetic $J_2$ as expected. The interlayer spin correlation $J_c$ is ferromagnetic in $x = 0.2$ and antiferromagnetic in $x = 0.5$.

Table S5. In-plane spin correlations obtained from the SPINVERT fits of $Sr_2CuTe_{1-x}W_xO_6$ diffuse magnetic scattering and the expected spin correlations for complete Néel or columnar magnetic order. $J_1$ corresponds to the side of the square, $J_2$ to the diagonal, $J_3$ to a Chess Knight move (two lengths along one side and one along a perpendicular side) and $J_c$ to the interlayer distance.

|        | $r_i$ (Å) | $Z_i$ | Néel | Columnar | $x = 0$   | $x = 0.2$ | $x = 0.5$ | $x = 1$   |
|--------|-----------|-------|------|----------|-----------|-----------|-----------|-----------|
| $J_1$  | 5.4       | 4     | -1   | 0        | -0.12(1)  | -0.12(1)  | -0.01(2)  | 0.01(1)   |
| $J_2$  | 7.6       | 4     | 1    | -1       | 0.02(1)   | 0.07(1)   | -0.12(1)  | -0.13(1)  |
| $J_c$  | 8.4       | 2     | 1    | -1       | 0.15(2)   | 0.05(2)   | -0.04(2)  | -0.09(2)  |
| $2J_1$ | 10.8      | 4     | 1    | 1        | -0.01(1)  | 0.02(1)   | 0.08(2)   | 0.03(2)   |
| $J_3$  | 12.1      | 8     | -1   | 0        | -0.01(1)  | -0.01(1)  | 0.02(1)   | 0.04(1)   |
| $2J_2$ | 15.3      | 4     | 1    | 1        | 0.0(1)    | -0.01(1)  | 0.05(2)   | 0.04(2)   |